# Temperature and force dependence of nanoscale electron transport via the Cu protein Azurin

*Wenjie Li,[†] Lior Sepunaru,[†,‡] Nadav Amdursky,[†,‡] Sidney R. Cohen[&], Israel Pecht,[§] Mordechai Sheves,[*,‡] and David Cahen[*,†]*

[†]Department of Materials & Interfaces, [‡]Department of Organic Chemistry, [&]Chemical Research Support and [§]Department of Immunology, Weizmann Institute of Science, Rehovot 76100, Israel

*Corresponding authors:  David Cahen         e-mail: david.cahen@weizmann.ac.il

Mordechai Sheves    e-mail: Mudi.Sheves@weizmann.ac.il

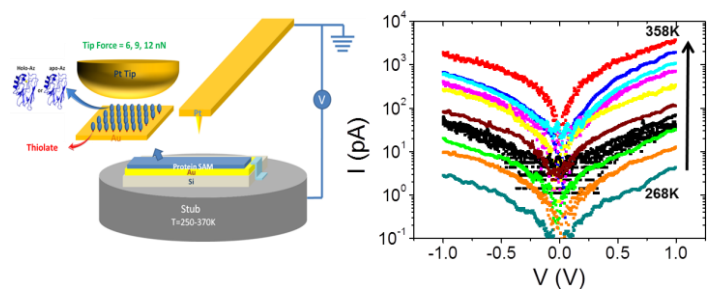

**TOC graphic:**

**Abstract**

The mechanisms of solid-state *electron transport (ETp)* via a monolayer of immobilized Azurin (Az) was examined by conducting probe atomic force microscopy (CP-AFM), both as function of temperature (248 - 373K) and of applied tip force (6-12 nN). By varying both temperature and force in CP-AFM, we find that the ETp mechanism can alter with a change in the force applied via the tip to the proteins. As the applied force increases, ETp via Az changes from temperature-independent to thermally activated at high temperatures. This is in contrast to the Cu-depleted form of Az (apo-Az), where increasing the applied force causes only small quantitative effects, that fit with a decrease in electrode spacing. At low force ETp via holo-Az is temperature-independent and thermally activated via apo-Az. This observation agrees with macroscopic-scale measurements, thus confirming that the difference in ETp dependence on temperature between holo- and apo-Az is an inherent one that may reflect a difference in rigidity between the two forms. An important implication of these results, which depend on CP-AFM measurements over a *significant* temperature range, is that for ETp measurements on floppy systems, such as proteins, the stress applied to the sample should be kept constant or, at least controlled during measurement.

Azurin (Az) is an electron-mediating protein, functional in the bacterial energy conversion system, which has been studied extensively, mostly by spectroscopic[1-3] and electrochemical measurements[4-10] in solution, as a model system for electron transfer (ET) via proteins. *Solid-state electron transport* (ETp) measurements have been reported on several types of proteins,[11-13] including Az, which has been investigated quite intensively, both by nano-scale techniques[14-19] and by macro-scale electrodes.[20-23] Within the latter context we reported recently macroscopic scale ETp measurements via Az monolayers,[21] which showed temperature-independent currents via holo-Az. ETp via the Cu-depleted form of Az, apo-Az, was found to be thermally-activated at T > 180K and temperature-independent at T < 180K.

Investigating these current-voltage measurements with nano-scale contacts, may scrutinize two important issues: (i) While the observed temperature independence of the currents through a 3.5 nm thick protein (see below) was quite remarkable, we sought a different measurement method, less likely to be influenced by possible defects (in the monolayer), to assure that this is indeed an intrinsic property of Az. (ii) Use of conducting probe atomic force microscopy, CP-AFM, allows to gain information on the ETp under different pressure applied to the protein. Up to now, only a simple connection between tunneling currents and the force, applied to an organic monolayer was found.[24, 25] With proteins (and likely also for other soft materials), increasing the applied force may cause structural changes, which should be considered, in addition to a simple compression that increases current flow. By combining temperature dependent current transport measurements and the unique AFM capability of force control on a nano-scale contact area, we can use ETp as an indicator for force-dependent structural changes in proteins. These force-dependent measurements can also provide comparison with macroscopic measurements, where only a negligible gravity force and a constant adhesive force (both from the contact pad) act on the proteins. While, in principle, force control in macroscopic measurement on a monolayer is possible, its use is hampered by the fact that even slight variation in monolayer uniformity may cause penetration of the protein monolayer by the electrode.

In a series of experiments, Davis and coworkers observed ETp through Az by CP-AFM.[14, 16-19] From fits of measured I-V curves to Simmons' non-resonant tunneling model they extracted values for barrier height and length of the Az monolayer at different applied tip forces (at room temperature). Along with ETp experiments via Az, current-voltage CP-AFM measurements of other protein monolayers in a solid state environment have been reported, and the data were fitted to different ETp models (for example Fowler-Nordheim tunneling[26]).[15, 27, 28] CP-AFM allows measuring ETp through a limited number of molecules, located between the tip and the surface, or attached to the tip, to the surface or to both. The effective radius of the AFM tip limits the contact area of the measurement and in the present work we estimate that we measure up to ~50 Az molecules. Smaller contact area increases the probability of resolving small defect-free areas for measurement. At the same time we sacrifice sensitivity (lower currents), compared to use of the much more sensitive macroscopic electrode measurements that average over a large area. The two approaches can, thus, be viewed as complementary ones.

According to Marcus theory,[29] $k_{ET}$, the rate constant for electron transfer, ET between a donor (D) and an acceptor (A), is given by:

$$k_{ET} \propto H^2_{D-A}(l_{D-A})\exp(-E_a/k_BT)$$

(1)

where $E_a$ is the activation energy of the process, $T$ *is* the absolute temperature, $k_B$ is Boltzmann's constant and $H_{D-A}$, is the degree of electronic coupling between D and A, which also depends on $l_{D-A}$ the separation distance between them. While in ET studies $l_{D-A}$ is the commonly varied parameter, and varying the temperature maybe problematic over a wide range (also in view of solvent presence), for ETp the opposite situation holds. The Az is sandwiched between two electrodes and, in our set-up is covalently attached to the bottom electrode by disulfide bridge through its cysteine residue (Cys3 or Cys26). This configuration introduces a defined and fixed distance between the two electrodes, $l$ =3.5 nm, although in CP-AFM measurements one can change the tip-surface distance by applying different probe forces.[30] Using molecular dynamics simulations, Davis and coworkers proposed that the secondary structure of the protein changes drastically upon applying increasing forces to the protein layer.[19] Accordingly, as will be shown and discussed below, at different probe forces, we can expect the medium that separates the tip and the surface to be affected. This poses a problem for certain nanoscopic measurements, such as those done by dynamic scanning tunneling microscopy (STM),[31] because the protein will be mechanically stressed and, possibly structurally affected, during most of the measurement. Here we report temperature- and force-dependent CP-AFM of holo- and apo-Az, as a monolayer, in a solid state-like junction over the temperature range of 248-373K and applied forces of 6-12 nN.

**RESULTS**

In the employed configuration, a smooth (r.m.s. roughness of ~0.7 nm over 4×4 μm scan) and continuous Au layer is deposited on an H-terminated Si substrate. A self-assembled monolayer of holo- or apo-Az was coupled covalently to the Au substrate via S-Au bonding between the Au surface and the relatively exposed cysteine thiolate of Az (Fig. 1). To confirm the establishment of the protein monolayer, topographic (amplitude) and phase images were measured in semi-contact mode with different amplitude reduction or amplitude set point ratio (97%, 94% and 91%, respectively) (Fig. 2). A lower set point ratio corresponds to a higher tip force on the surface. The relative influence of attractive (adhesive) forces changes as the set point ratio is lowered, particularly on compliant surfaces, leading to modification of topography and phase images.[32-34] As we lowered the set point ratio, i.e., increased the tip force, we clearly observed a change in the topography images, and corresponding changes in the phase images. The observed changes in the topography and phase images with different tip forces suggest the presence of a soft layer, i.e. proteins. To assess the thickness of the protein monolayer, a square area was scanned in the contact mode with a large feedback force (160 nN). The applied force is sufficiently large to scratch away the monolayer. Following the contact mode scratching procedure we reverted to semi-contact mode to re-scan over a larger area, centered around the resulting trough (Supporting Information, Figure S1 a). The obtained image demonstrates that the probe removed the protein over a square area. The cross-section (Supporting Information, Figure S1 b) shows that the thickness of the protein layer is about 3.5 nm, the length of Az established by its three-dimensional structure determination.

The measured I-V of holo- and apo-Az as a function of applied force (conducted at room temperature), is shown in Figure 3. The currents obtained for both proteins increased with probe force. For the holo-Az junctions at 0.5 V bias, the currents ranged between 0.1-1 nA for tip forces of 6 – 15 nN, in agreement with the values reported by Zhao and Davis.[14, 19] Clear shorts were observed everywhere with > 40 nN tip force, suggesting the tip pinches through the protein layer at this force. The currents across apo-Az junctions spanned a range of 0.002 – 2 nA for the same force variation. The higher sensitivity of apo-Az to the applied tip force suggests that it is more flexible than holo-Az. Indeed, the crystal structure of apo-Az, though very similar to that of holo-Az, indicates more flexibility,[35-37] particularly at the, for apo-Az empty, metal binding site. If a low force is applied at room temperature, the observed current across apo-Az junctions at 1 V was an order of magnitude lower than that across the holo-Az junctions (Fig 3, black curve). This difference resembles that obtained in our macroscopic measurements[21] and is qualitatively consistent with results of conduction measurements via another metal-containing protein (Ferritin) by CP-AFM[27], STM[38] or, for Az in a three-terminal configuration.[39] This decrease in current magnitude highlights the role of the Cu ion as an efficient ET and ETp mediator in Az.

Figure 4 shows the current-voltage characteristics of holo-Az (Figs. 4a,b) and the natural logarithm of the currents for voltage sweeps from -1 V to +1 V for apo-Az junctions (Fig. 4c,d) *as a function of temperature*. The I-V curves were measured over more than 30 random spots for each temperature at constant probe force (6 nN). For each temperature we took the average of all *I-V* curves, excluding those exhibiting shorting or insulating behavior (~20%) The distribution of current intensities was normal (inset of Fig. 4a); average currents were used for further analysis. The *I-V* measurements were performed at each temperature only after the system had remained stable for at least 30 minutes. As can be seen in figure 4, holo-Az and apo-Az exhibit distinct ETp characteristics as a function of temperature. The *I-V* curves of the holo-Az junctions remained essentially invariant with temperature from 248 to ~370 (Fig. 4a). The measured current via the junctions at +0.5 V and – 0.5 V as a function of the inverse temperature (Fig. 4b) clearly demonstrates a temperature-independent ETp process across holo-Az. Thus, the CP-AFM results, obtained with a relatively low tip force of 6 nN, confirm our macroscopic ETp observations.[21] In the CP-AFM measurements, we observed a constant adhesion force of 5 nN, which was measured as the snap-out force during retraction of the AFM tip. Thus, the total load applied on the proteins is 11 nN. In the macroscopic measurement using an Au pad, the force, exerted by the Au on the proteins, is mainly governed by the intrinsic adhesion force. Symmetry considerations imply that this force is similar to that of this intrinsic metal tip/protein adhesion. Since in the macroscopic scale measurements there is no significant additional external load, the effective pressure applied to each protein molecule is less than half that in the nano-scale measurements.

The role of the Cu redox center in the ETp process across the protein was measured on a monolayer of apo-Az. Similar (optical) heights were deduced from ellipsometry for apo-Az and holo-Az (18Å), comparable to the observed height on Si surfaces[20], corresponding to similar packing of the monolayer on the Au surface. The junctions of apo-Az were prepared, similarly to those of holo-Az, and the same AFM tips were used to measure the *I-V* characteristics of the two proteins. Figure 4c shows the *I-V* curves of the apo-Az junction on a logarithmic current scale at different temperatures, in the range of 248-368K. Figure 4d shows the logarithm of the currents through the junctions at +0.5 V as a function of inverse temperature. A linear dependence is clearly observed. Thus, removing the Cu ion changes the dominant mechanism of ETp across the protein from temperature-independent to the more common thermally activated type, in agreement with the macroscopic ETp results.[21]

Upon raising the temperature further we observed a sharp irreversible decrease of the currents for apo- and holo-Az junctions (marked in circle in Fig. 4b and d), this is in contrary to the reversible temperature dependent that we observe (by heating and cooling) in the monolayer junction at lower temperatures. The most likely reason for the irreversible decrease in the currents is the denaturation of the protein. The reported $T_m$ values of apo- and holo-Az in solution are 335K and 355K respectively,[40] which are slightly lower than the observed denaturation in the macro- and nano-scale measurement. Higher stability, consequently leading to higher $T_m$ has also been observed previously in other dry proteins.[41] In our macroscopic scale experiments these current drops appear at 350K and 360K for apo and holo-Az, respectively.[21] The discrepancies of 5-15K may be due to the different coupling modes to the electrodes, which may affect the stability of the protein as well.

The effect of applied tip force on the current as a function of temperature is shown in Figures 5a and b, which give the currents at +0.5 V applied bias on the tip as a function of inverse temperatures for holo- and apo-Az, respectively, at 6, 9 and 12 nN tip force. As shown in Fig. 3 for room temperature currents, increasing the force, leads to higher current. However, we find a distinct difference in the temperature dependence of the current between apo- and holo-Az, at different tip forces. While currents via apo-Az increase with increasing tip force, as expected from a decrease in the tip-substrate distance, their temperature dependence remains similar. By fitting the curves in Fig. 5b to the Arrhenius equation, we can estimate the ETp activation energy, $E_a$, of apo-Az to be 600±100 meV at all of the applied forces. In contrast, the currents via holo-Az change from temperature-**in**dependent at 6 nN to thermally activated at higher applied forces (9 and 12 nN) at temperatures > 310K. Below 310K the currents through holo-Az remain temperature-independent at all of the applied forces, but currents are higher for higher applied force, which can be the result of a decrease in tip-substrate distance. These differences between apo- and holo-Az can also be seen in a plot of the measured currents as a function of the applied tip force (Fig. 6), obtained by combining simultaneous force and current traces as a function of time by high speed data acquisition in Peak Force TUNA$^{TM}$ mode. Our results show that while the slopes of the currents as a function of the tip force for apo-Az (Fig. 6a) remain similar at different temperatures, the slope for holo-Az increased significantly with temperature.

The above mentioned irreversible drop in current upon heating was observed at similar temperatures for both holo-Az and apo-Az, with 6, 9 and 12 nN tip force applied. This similarity suggests that any structure distortion of the protein caused by a tip force of up to 12 nN is minor, compared to the change involved in denaturation.

## DISCUSSION

***Comparison of current densities observed in macroscopic and nanoscopic contact measurements***

Thermally activated ETp was previously observed in CP-AFM studies via > 4 nm long conjugated molecules (with temperature-independent behavior for shorter molecules).[42, 43] The Arrhenius plots of the currents as a function of inverse temperature observed in those experiments were interpreted in terms of a hopping ETp mechanism. This may well be the dominant ETp mechanism via apo-Az. While Az, with or without its Cu redox centre, is not a conjugated system, apo-Az behaves *qualitatively* in a similar fashion to conjugated molecules ("molecular wires"), but with measured currents of 10's of pA, rather than the μA currents that flow through conjugated molecules of similar molecular length at 0.5 V applied bias.[43] The contact area for these conjugated molecules was estimated at 50 nm$^2$.[43] We can estimate our contact area by using the following relation derived from Herzian contact mechanics:[24]

$$A = \pi \left( \frac{3 P_{eff}}{4 E^*} r \right)^{2/3}$$

where $r$ is the tip radius, $P_{eff}$ is the load and $E^*$ is the effective modulus, which we estimate to be ~100 MPa.[44, 45] From this calculation we assess the contact area in our studies to be ~450 nm$^2$. After normalization of the contact area differences, we find the current densities for both holo- and apo-Az to be 5-6 orders of magnitude lower than those of conjugated molecules. In comparison, an insulating layer of alkyl chains of similar thickness, at comparable force and tip radius should pass $10^{-18}$A (extrapolating from CP-AFM measurement on $CH_3(CH_2)_{11}SH$, ~500 pA at 0.5 V, and using β=1.1 A$^{-1}$ [25]). This observation by itself makes proteins, in terms of conductivity, more akin to molecular wires than to insulators.

In general, the CP-AFM current-temperature behavior of Az with nano-scale electrode contact is consistent with our previously reported macroscopic results,[21] in that they both show temperature-independent ETp of holo-Az and thermally activated ETp of apo-Az. There are however some differences between the CP-AFM and the macroscopic conductance measurements[25] in the normalized current densities across the proteins and in the activation energy, $E_a$, of apo-Az that was derived from the data.

*Current Densities*

From the nano-scale measurements (AFM tip radius of 20 nm) we calculate for holo-Az junctions a current density of ~100 A/cm$^2$ at 1 V, while in the macroscopic measurements (contact area of 0.2 mm$^2$) it was ~3x10$^{-3}$ A/cm$^2$, namely five orders of magnitude lower. We ascribe this difference to differences in the way the proteins are immobilized between the two electrodes in the two measurements. In the macroscopic conductance measurements the substrate was a p$^{++}$-Si surface with a 1 nm thick layer of silicon oxide, with, as linker, a self-assembled monolayer of a C$_3$ organo-silane on top of it. The linker molecules were covalently bound to the Az proteins via a disulphide bond.[20] The top electrode was a macroscopic pad of gold (~0.2 mm$^2$), deposited by the lift-off, float-on (LOFO) technique,[46] to contact the protein layer. Thus, in the macroscopic measurements the two contacts were separated not only by the protein monolayer, but also by two additional insulating layers, silicon oxide and organic linker, with a combined thickness of ~16 Å). These two additional insulating layers will decrease the currents by some five orders of magnitude, assuming a current decay factor (for insulating molecular layers) of β = 0.7 A$^{-1}$.[12]

*Thermal activation energy for ETp via apo-Az*

The distinct experimental setups may also be the cause for the different calculated ETp activation energies of apo-Az, i.e., 320 meV from the macroscopic measurements[30] and ~600 meV from the CP-AFM measurements. Because the holo-Az results exhibited the same temperature-independent behavior in the two different experimental setups, it is very unlikely that the different electrodes are the cause for this difference in activation energy. A possible reason is that the covalent S-S bond between one of the two exposed cysteines and SH group of the linker in the macroscopic measurements is stronger, than the Au-S bond that binds the protein to the Au substrate in the CP-AFM measurements. The smaller coupling to the electrode might allow additional vibration modes at the Au/protein interface, which are hindered in the case of the disulfide bridge bond that holds the protein in the macroscopic configuration.

*Tip pressure dependence of the ETp*

A striking result of this study is the observed distinct temperature dependence of the current through holo-Az as a function of applied force (Figs. 5 and 6). As discussed previously, the higher sensitivity to the applied tip force of currents through apo-Az (at room temperature) than through holo-Az, (Fig. 3) is interpreted as being a result of the increased flexibility of apo-Az, compared to holo-Az. This increased flexibility of apo-Az causes the tip-surface separation distance to be smaller at high forces than in the holo-Az surface, which explains the larger increase in currents with increasing force via apo-Az. In other words, Young's modulus of apo-Az protein is lower than holo-Az, leading to a more significant change in the separation distance between the tip and the gold substrate, due to protein compression, hence leading to more pronounced changes in currents between the electrodes.

An increase in the applied tip force on top of the holo-Az monolayer results in the AFM tip pushing toward the Cu binding site (which is located on the tip-side of the junction as shown in the scheme of Fig. 1). Marshall et al.[47] have shown that minor alterations of the hydrogen-bonding network in the second coordination sphere of the Cu site may cause pronounced changes in the ET properties of holo-Az, which consequently can change its flexibility.[48-50] The change (at T > 310 K) from temperature-independent to thermally-activated ETp via holo-Az suggests a relation between the protein structure and flexibility, and the mechanism of ETp through it.[51] The suggested correlation between flexibility and conductivity[47] has been indicated by measurements of ETp via peptide nucleic acid monolayers, where attenuation in currents was observed for structures comprising the same sequence, with and without methylation, a modification that increases the structural rigidity.[52] Therefore, we suggest that the protein's flexibility (in the case of holo-Az), together with the temperature of the system (for apo-Az) are dominant parameters that control both the mechanism and the efficiency of ETp via the protein.

The impact of applied forces on the ETp characteristics of proteins suggests that the compressive and tensile stress applied to the examined protein need to be taken into consideration when its electrical conduction properties are investigated. , In similar ETp measurements that were conducted via alkyl chains, currents began to increase only at > 12nN tip force,[25] suggesting a more rigid structure than that of proteins. Furthermore, the force applied by the tip on an alkyl chain has apparently a simpler effect, than that on proteins, where more complex secondary structures can be affected. Thus, consideration of the applied force appears to be particularly relevant for nanoscopic approaches, such as STM and CP-AFM, for measurements of proteins, as shown here.

**CONCLUSIONS**

CP-AFM measurements of protein monolayers show that, at low tip-force, ETp through holo-Az is temperature-independent over a significant range of temperatures (248-373K), while for apo-Az the process is thermally activated. This observed difference is in line with our earlier macroscopic ETp results. As the tip-force increases, ETp through holo-Az changes from temperature-independent to thermally activated at higher temperatures (> 310K). It is likely that the mechanism of ETp through apo-Az does not change with pressure, as the thermal activation energy does not change. The currents, though, increase with increasing tip force, an effect that can by ascribed to decreases of through-space ETp gaps (by compression of tunneling pathways) in the protein, consistent with results of Meier et al.[53] The results obtained with holo-Az do suggest a pressure-induced change in ETp mechanism at temperatures >310K, which may be due to a pressure-induced change in protein structure. Even though this change is likely to be minor, because it does not significantly alter the denaturation temperature, it does manifest itself in ETp. Thus, this result illustrates the remarkable sensitivity of ETp to both relatively minor structural changes, as well as to major changes in the protein (denaturation). The strong relation between protein flexibility and applied force, and its impact on the temperature-dependence of ETp via Az, indicates the importance of the applied forces for current measurements in scanning microscopy configurations (cf., review on biomolecule conductance[11] and report on force-dependent conductivity[14]).

Finally, our observation directly points to a general issue that is worth considering: force-mediated effects in nano-scale electrical measurements. The applied forces should be the lowest that still assure reproducible ETp. Because a change in applied force such as the one demonstrated here, can affect ETp, changes in force during ETp measurements, should be avoided or their effect should be considered. This consideration is especially important for ETp measurements through proteins because of the relative ease by which they can change conformation, compared to more rigid molecules, such as, e.g., conjugated or alkyl ones. The combination of controlled temperature and environment, together with a defined (and reported) tip diameter and force are critical parameters that define the initial conditions of the system, and, hence, the output of the measurement.

## METHODS

**Preparation of the substrates.** Silicon wafers (p type, boron doped, <100> single side polished, $\rho <$ 0.001 $\Omega \cdot cm$), were sonicated for 2 min with ethyl acetate, acetone and ethanol. Immediately after that, the wafer was etched for 1 min with 2% HF, washed with Milli-Q (18M$\Omega$) water and cleaned again with fresh piranha solution for 20 min (7/3 v/v of $H_2SO_4/H_2O_2$) at 80°. After cleaning with piranha the wafers were rinsed thoroughly with Mili-Q and etched with 2%HF again, resulting in a Si-H surface termination. The wafers were immediately stored in a container, filled with nitrogen. The samples were loaded into an e-beam evaporator and, when a vacuum of $< 5*10^{-6}$ mbar was reached, 2 nm of Cr (serving as an adhesion layer) followed by 50 nm of Au were evaporated with a deposition rate of 1Å/s. The gold-coated Si samples were then cut into 1*1 $cm^2$ slides, cleaned for 10 min with UV/ozone treatment followed by a 30 min immersion in ethanol.[54]

**Preparation and characterization of the proteins.** Azurin was isolated from *Alcaligenes faecalis* by the method of Ambler and Wynn.[55] Apo-Az solution was prepared as described.[20] Holo-Az and apo-Az monolayers were prepared by immersing the gold substrates in a ~1 mg/mL solution of azurin in 50 mM ammonium acetate ($NH_4Ac$) buffer (pH 4.6) for 2 h followed by rinsing in clean $H_2O$ and drying under a fine nitrogen stream. The protein monolayers were characterized by ellipsometry that yielded 1.8 nm optical thickness, which corresponds to values obtained in our previous studies of Az on Si surfaces.[20] The actual thickness is 3.5 nm, as found by AFM (Supp. Info., Fig. S1b)

**AFM imaging.** The topography of the self-assembled monolayer of proteins was characterized by AFM in semi contact mode under $N_2$ purge. A Solver P47 SPM system (ND-MDT, Zelenograd Russia) and Pt coated Si probes (NSC36, 75kHz, 0.6 N/m, MIKROMASCH) were used. The topography images and the phase images were taken simultaneously at a scan rate of 1 Hz. The applied force from the tip on the proteins, during the topography imaging, is at least one order of magnitude smaller than those used for CP-AFM measurements. (supporting information)

**CP-AFM measurements as a function of temperature.** The CP-AFM measurements *as a function of temperature* were performed with a Multimode/Nanoscope V system (Bruker-Nano, Santa Barbara, CA USA) under constant $N_2$ flow purge. All-metal Pt AFM probes with nominal force constant of 0.8 N/m (25PT300B, Rocky Mountain Nanotechnology, Salt Lake City Utah USA) and tip radius of 20 nm, were used for the CP-AFM measurements. The probes were brought into contact with the self-assembled protein monolayer using a constant force feedback (contact mode). Given the tip radius, ~50 proteins were measured and averaged, so our temperature and force dependent measurements are less affected by the drifting and penetration of the proteins than in the case of a single protein. A tip force of 6 nN was used to avoid changes in the protein, following ref.[19] Temperature was controlled by a Bruker heater/cooling system that uses an inert atmosphere flow, rather than a closed vacuum system as in the case of our macroscopic measurements. With this set-up we can obtain current voltage (*I-V*) curves between 248K and 373K. At temperatures below 273K (0º C) the protein surface was monitored constantly to verify that no ice is formed, ascertaining the low moisture level in the system. For each temperature, the *I-V* measurements were performed only after the system had stabilized and remained stable for at least 30 minutes. 30 *I-V* curves over an area of 1×1 μm$^2$ were taken and averaged for each temperature. The drifting of the tip position will not affect our measurements. Raw data of *I-V* curves from proteins samples at room temperature were plotted in Figures S2 (Supporting Information) to show the variation of the *I-V* curves. Whenever drastic current drops occurred, the tip was cleaned, using a high voltage pulse (-10 to+10 V in 0.1 sec). The tip force was constant during the measurements.

**AFM Peak Force TUNA Mode.** An extension module for current measurement was used to enable the current mapping under AFM Peak Force (PF-TUNA$^{TM}$) mode. In Peak Force TUNA mode, the probe was cycled in and out of contact with the surface at 1 or 2 kHz, while the tip is scanned across the sample at a rate of 1 Hz per scan line. The fast data acquisition, coupled with feedback loop control, the maximum force on the tip for each individual cycle. Current, force and other mechanical properties are recorded during these controlled tip-sample contact cycles. Peak current is recorded at maximum force applied to the sample (Peak Force set point). Measurements were performed in both imaging mode and spectroscopy mode. In the imaging mode, a current map is acquired simultaneously with topography at a tip bias of +0.5 V with a tip Peak Force of 6, 9 nN and 12 nN. The currents from the image were averaged for each such mapping at a given temperature. In spectroscopy mode, the tip was ramped into and out of the surface at high cycling frequency (1 kHz), while recording simultaneously both force and current as function of time. The current force curve was plotted by combing the two simultaneous traces.

**Acknowledgments**  WL thanks Daniel Frisbie and Liang Luo (Un. Minnesota) for helpful guidance with, and discussions on CP-AFM measurement procedures. LS thanks the Eshkol program for financial support. NA thanks the Clore program for financial support. We thank the Minerva Foundation (Munich), the Kimmelman center for Biomolecular Structure and Assembly, the Kimmel centre for Nanoscale Science and the Grand Centre for Sensors and Security for partial support. MS holds the Katzir-Makineni chair in Chemistry. DC holds the Schaefer Chair in Energy Research.

*Supporting Information Available:* additional information about AFM imaging, AFM tip scratching procedure, -protein thickness measurements and raw data of I-V curves for Az and apo-Az at room temperature. This material is available free of charge via the Internet at http://pubs.acs.org.

**Figure Captions**

**Figure 1.** Schematic (not to scale) of junction configuration employed in the ETp measurement via the proteins. (PDB ID: 1AZU)

**Figure 2.** AFM topography (left) and phase (right) images of holo-Az with different amplitude set point ratios (97%, 94% and 91%, respectively), corresponding to different forces applied on the proteins.

**Figure 3.** Representative force-dependent I-V curves of (a) holo-Az and (b) apo-Az at room temperature. The current increases with force from 6 nN to 15 nN at 1 V are 9.7 ± 2 times and 5.5 10-2 ± 3.$10^2$ times for holo-Az and apo-Az, respectively. See Suppl. Info. for a raw data set.

**Figure 4.** (a) Averaged I-V curves of holo-Az at temperatures from 248 to 373 K. The inset shows a normal distribution of the currents taken at +1 V. (b) Currents of holo-Az at +0.5 V (black squares) and –0.5 V (red triangles) as a function of inverse temperature. (c) Semi-logarithmic averaged I-V curves of apo-Az at temperatures from 268 to 368 K. (d) Currents of apo-Az at +0.5 V as a function of inverse temperature. The black curves in (a) and (c) are for the samples after denaturation. They correspond to the circled points in (b) and (d), respectively. Error bars were calculated by the standard deviation of I-V curve measurements for each temperature.

**Figure 5.** ETp temperature dependence for (a) holo-Az and (b) apo-Az junctions, shown as plots of current at 0.5 V (logarithmic scale) vs. inverse temperature at 6, 9 and 12 nN applied tip force. Error bars are based on the standard deviation of the "Peak Force" currents from current mapping.

**Figure 6.** Representative single measurement, current-force curves (logarithmic scale) at 0.5 V at different temperatures for (a) apo-Az and (b) holo-Az junctions. The slopes of the logarithmic current vs. force plots are 0.6 ± 0.1 for apo-Az for all temperatures, and 0.4 ± 0.1 at 288K and 0.9 ± 0.2 at 338 K for holo-Az. The currents at 288 K at the lowest forces reflect the sensitivity limit for single measurements (10 pA).

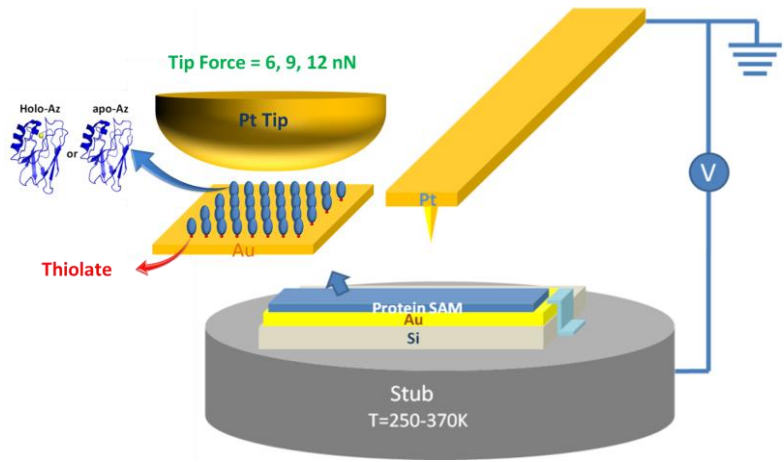

**Figure 1**

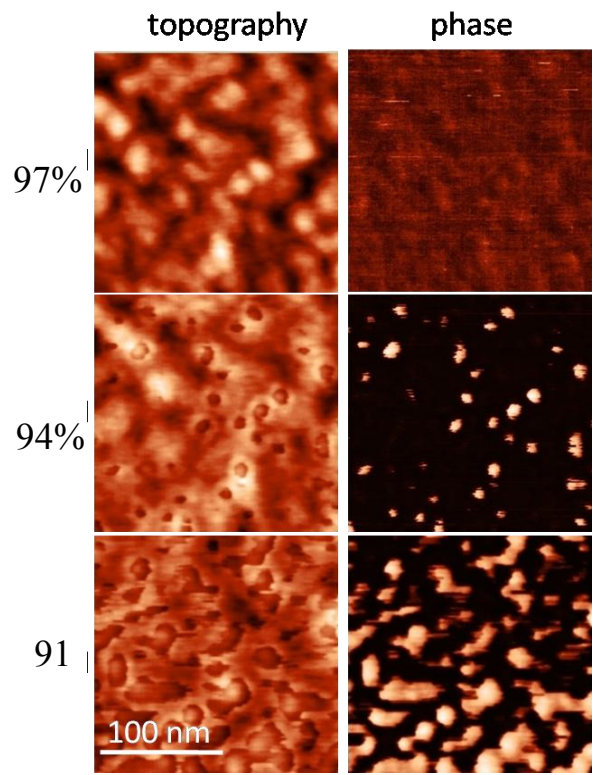

**Figure 2**

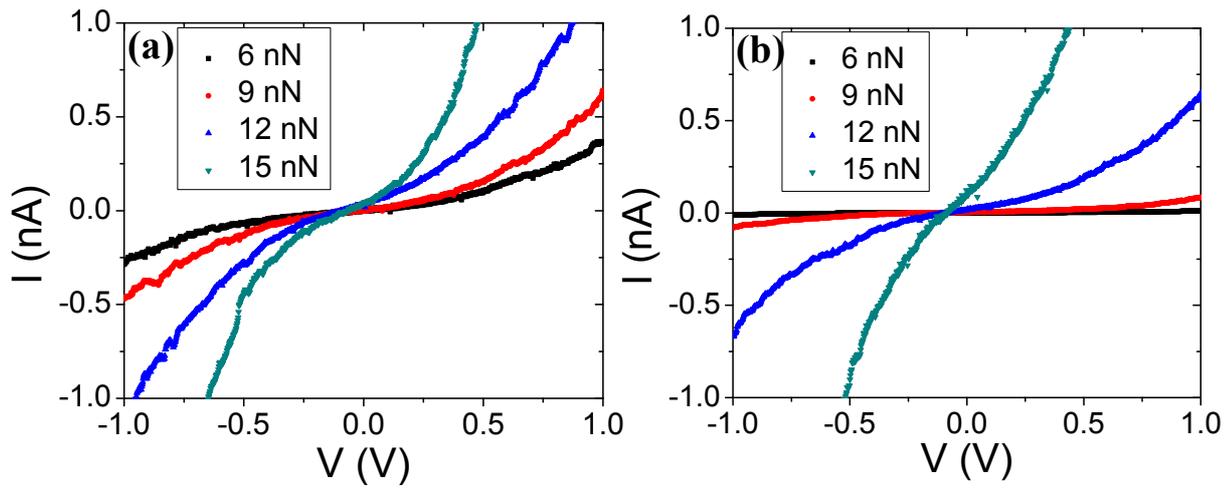

**Figure 3**

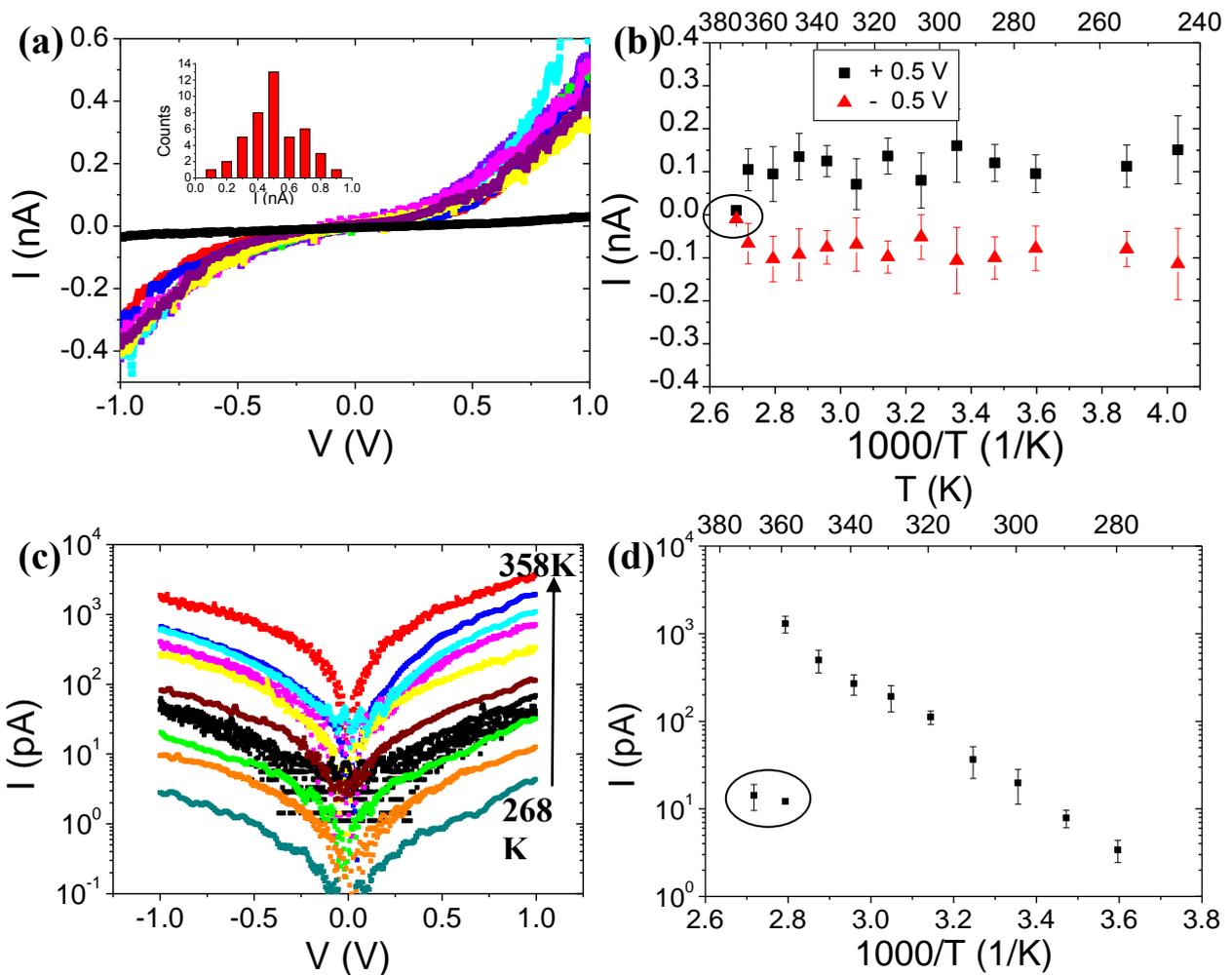

**Figure 4**

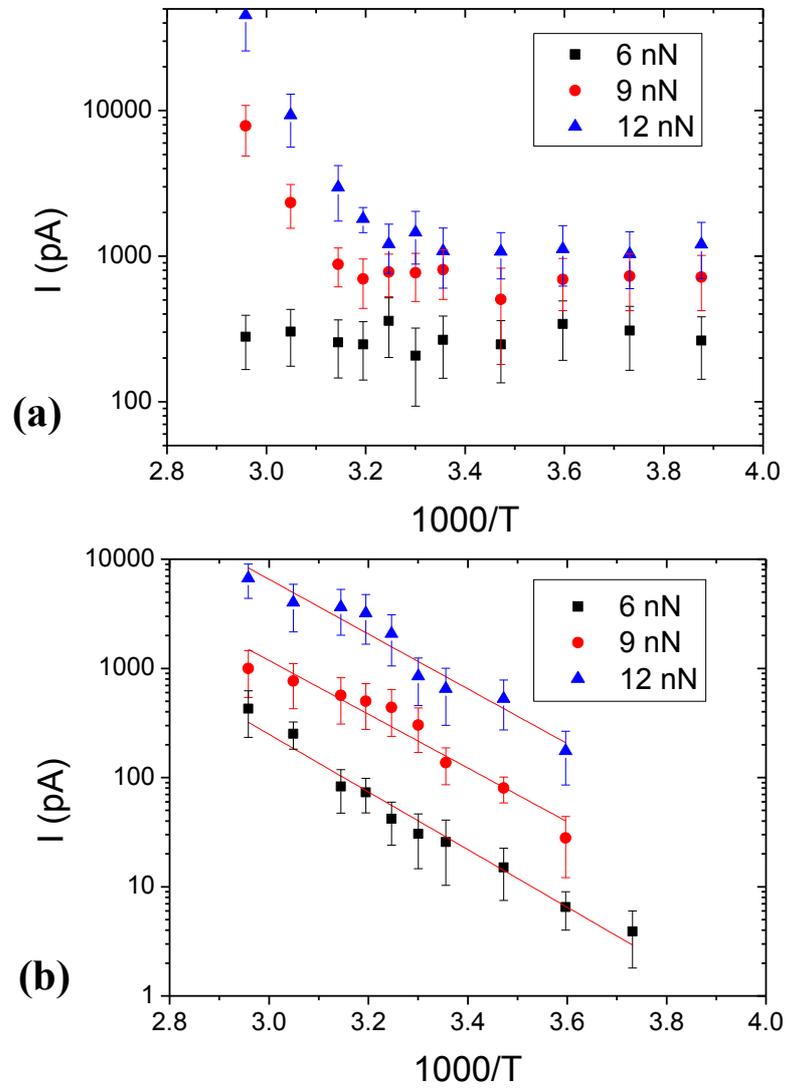

**Figure 5**

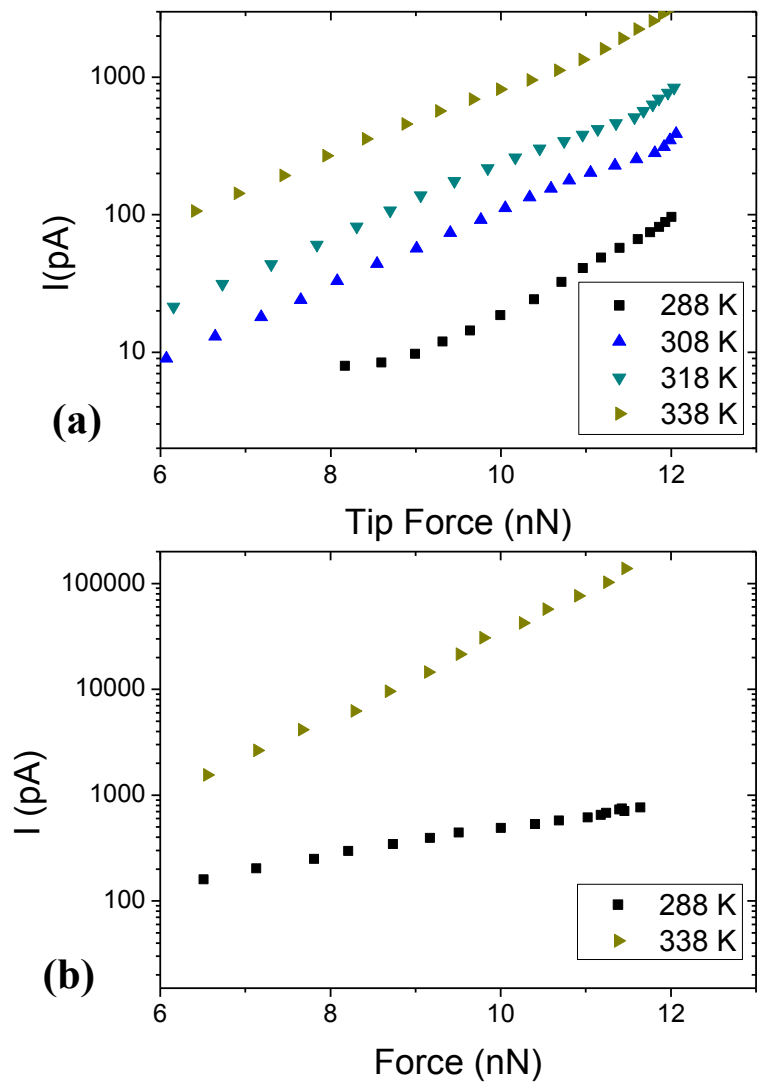

**Figure 6**

# Supporting Information

# Temperature and force dependence of nanoscale electron transport via the Cu protein Azurin

*Wenjie Li,[†] Lior Sepunaru,[†,‡] Nadav Amdursky,[†,‡] Sidney R. Cohen[&], Israel Pecht,[§] Mordechai Sheves,[*,‡] and David Cahen[*,†]*

[†]Department of Materials & Interfaces, [‡]Department of Organic Chemistry, [&]Chemical Research Support and [§]Department of Immunology, Weizmann Institute of Science, Rehovot 76100, Israel

*Corresponding authors:    David Cahen         e-mail: david.cahen@weizmann.ac.il
                          Mordechai Sheves    e-mail: Mudi.Sheves@weizmann.ac.il

**AFM imaging**

The topography of the self-assembled monolayer of proteins was characterized by AFM in semi-contact mode under $N_2$ purge. During a tapping cycle the tip spends only a small fraction of the cycle in repulsive contact with the surface. Therefore, the averaged force per cycle is quite small, as is the total energy imparted. An additional advantage to the semi-contact mode for obtaining clear images is that the tip detaches from the surface many times at each pixel, so that there is no shear force applied, which could lead to sample damage.

The much smaller force that can be used in the tapping mode than in CP-AFM is due to the different forces that are required for reproducible, stable mechanical or electronic contacts. The differences arise, because different physical interactions are involved. Quantitative calculation of the applied force is a complicated function of surface and probe properties, as well as specific operating conditions. Notwithstanding this uncertainty, with proper tuning of operating conditions, the forces applied during topographical imaging are at least an order of magnitude smaller than those used for CP-AFM scans.[1]

**Contact mode scratching procedure**

The scratching procedure was performed in a Solver P47 SPM system (ND-MDT, Zelenograd Russia). The operation mode of the AFM was temporarily switched to the contact mode with a programmed script and a 1x1 μm² square area was scanned with a large tip force (160 nN). The applied force is sufficiently large to scratch away the monolayer, but not sufficient to scratch the substrate surface. Following the contact mode scratching procedure we switched back to semi-contact mode to re-scan over a larger area, centered around the resulting hollow (Figure S1 a).

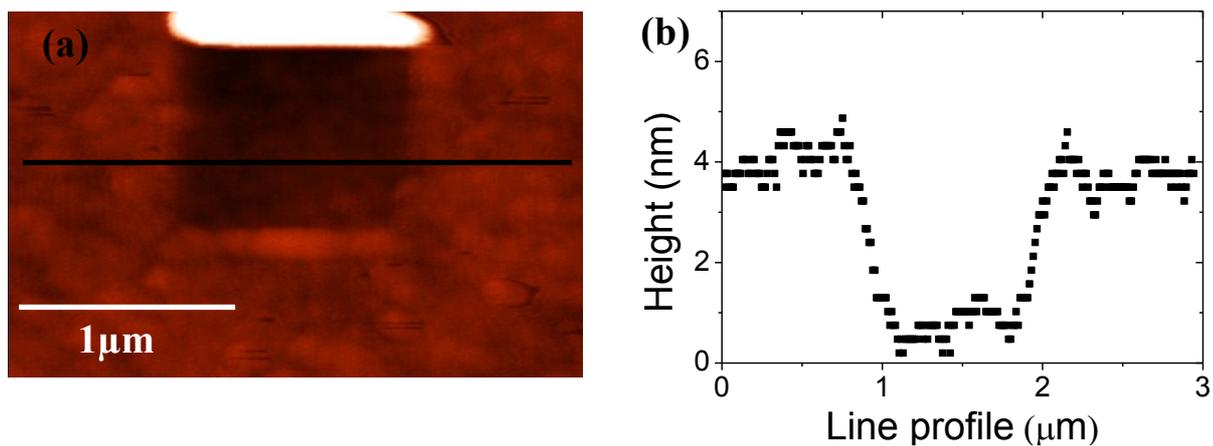

**Figure S1.** (a) AFM topography and (b) line profile of patterned Az monolayer from which part was removed by the AFM probe, enabling an estimate of the monolayer thickness (b). The island at the top center of (a) is the deposit of the proteins which were removed from the hollow in the image.

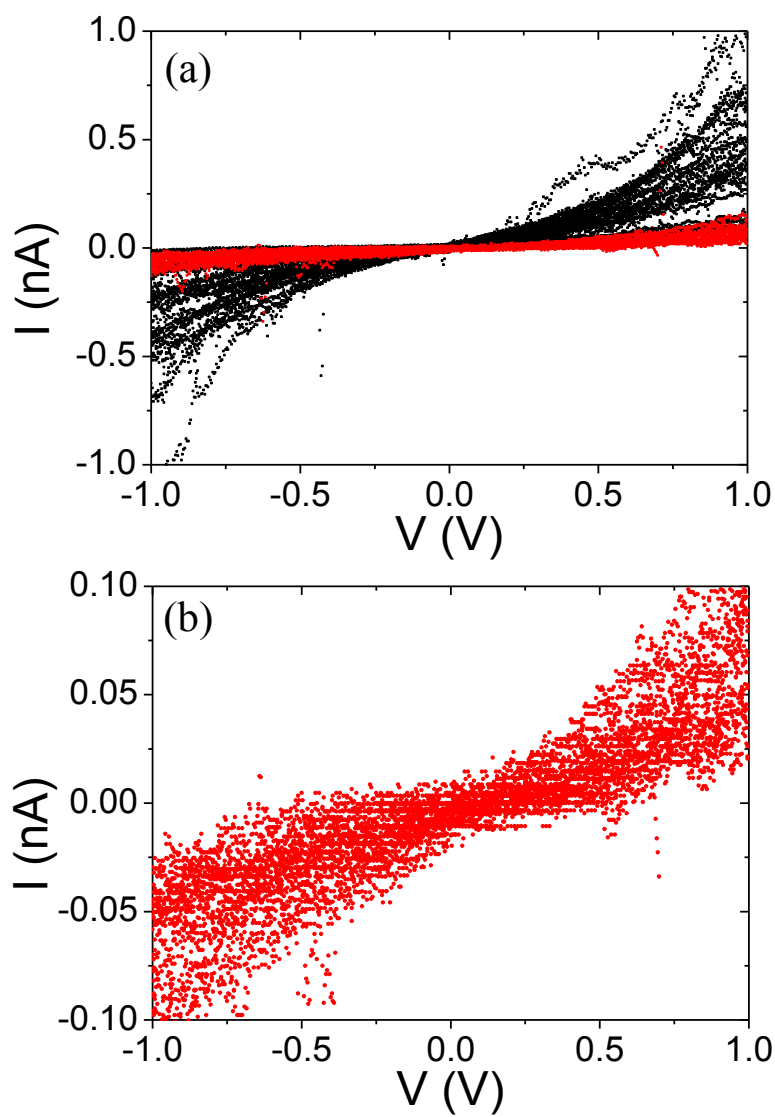

**Figure S2.** (a) Raw data of I-V curves for junctions with holo-Az (black squares) and apo-Az (red dots), measured at room temperature. (b) The I-V curves for the junctions with apo-Az of (a), but with expanded current scale. Tip force: 6 nN.

**Reference**

1. Garcia, R., and Perez, R., Dynamic atomic force microscopy methods, Surf. Sci. Rep., **2002**, Vol, 47, pp 197-301